\documentclass[superscriptaddress,nobibnotes,amsmath,amssymb,notitlepage,twocolumn]{revtex4-1}

\usepackage{pdfpages}
\usepackage{etoolbox} 

\usepackage{mathptmx,bm,physics,mathtools}
\usepackage{graphicx,hyperref,color} 
\hypersetup{colorlinks=true, linkcolor=blue, citecolor=blue, urlcolor=blue} 
\usepackage{dcolumn,multirow}

\graphicspath{{./img/}}

\newcommand{\beq}[1]{\begin{equation}\label{#1}}
\newcommand{\eep}{\;.\end{equation}}
\newcommand{\eec}{\;,\end{equation}}
\newcommand{\eeq}{\end{equation}}

\newcommand{\lb}{\left(}
\newcommand{\rb}{\right)}

\renewcommand{\th}{\theta}
\newcommand{\s}{\sigma}
\newcommand{\p}{\phi}

\newcommand{\D}{\Delta}
\newcommand{\G}{\Gamma}

\DeclareMathAlphabet{\mathcal}{OMS}{cmsy}{m}{n}

\newcommand{\C}{\mathcal{C}}

\renewcommand{\H}{\mathcal{H}}
\newcommand{\T}{\mathcal{T}}

\newcommand{\M}{\mathcal{M}}

\newcommand{\bvec}[1]{\mathbf{#1}}

\newcommand{\kv}{\bvec{k}}

\newcommand{\aM}{a_{\rm sc}}

\newcommand{\av}{\bvec{a}}
\newcommand{\bv}{\bvec{b}}
\newcommand{\rv}{\bvec{r}}

\newcommand{\kp}{\kv\cdot\bvec{p}}

\newcommand{\HarvardPhysics}{Department of Physics, Harvard University, Cambridge, Massachusetts 02138, USA}
\newcommand{\HarvardSeas}{John A.~Paulson School of Engineering and Applied Sciences, Harvard University, Cambridge, Massachusetts 02138, USA}
\newcommand{\ENSphys}{Département de Physique, École Normale Supérieure Université PSL, Paris 75005, France}

\makeatletter
\patchcmd{\@outputpage@head}{\@ifx{\LS@rot\@undefined}{}{\LS@rot}}{}{}{}
\makeatother

\begin{document}
\title{Twisted bilayer graphene revisited: minimal two-band model for low-energy bands}

\author{Daniel Bennett}
\affiliation{\HarvardSeas}

\author{Daniel T.~Larson}
\affiliation{\HarvardPhysics}

\author{Louis Sharma}
\affiliation{\HarvardSeas}
\affiliation{\ENSphys}

\author{Stephen Carr}
\affiliation{Department of Physics, Brown University, Providence, Rhode Island 02912-1843, USA}
\affiliation{Brown Theoretical Physics Center, Brown University, Providence, Rhode Island 02912-1843, USA}

\author{Efthimios Kaxiras} \email{kaxiras@physics.harvard.edu}
\affiliation{\HarvardSeas}
\affiliation{\HarvardPhysics}

\begin{abstract}
An accurate description of the low-energy electronic bands in twisted bilayer graphene (tBLG) is of great interest due to their relation to correlated electron phases such as superconductivity and Mott-insulator behavior at half-filling.
The paradigmatic model of Bistritzer and MacDonald [PNAS {\bf 108}, 12233 (2011)], based on the moir\'e pattern formed by tBLG, predicts the existence of ``magic angles'' at which the Fermi velocity of the low-energy bands goes to zero, and the bands themselves become dispersionless. 
Here, we reexamine the low-energy bands of tBLG from the {\em ab initio} electronic structure perspective, motivated by features related to the atomic relaxation in the moir\'e pattern, namely circular regions of AA stacking, triangular regions of AB/BA stacking and domain walls separating the latter. 
We find that the bands are never perfectly flat and the Fermi velocity never vanishes, but rather a ``magic range'' exists where the lower band becomes extremely flat and the Fermi velocity attains a non-zero minimum value. 
We propose a simple $(2+2)$-band model, comprised of two different pairs of orbitals, both on a honeycomb lattice: the first pair represents the low-energy bands with high localization at the AA sites, while the second pair represents highly dispersive bands associated with domain-wall states.  
This model gives an accurate description of the low-energy bands with few (13) parameters which are physically motivated and vary smoothly in the magic range. 
In addition, we derive an effective two-band hamiltonian which also gives an accurate description of the low-energy bands. 
This minimal two-band model affords a connection to a Hubbard-like description of the occupancy of sub-bands and can be used a basis for exploring correlated states. 
\end{abstract}

\maketitle


\section{Introduction
\label{sec:intro}
}

The physics of twisted bilayer graphene (tBLG) has proven remarkably rich, owing to the complexity of the moir\'e patterns formed for small twist angles in the range of $1^{\circ}$ or smaller.
The discovery of superconductivity and correlated electron behavior in this system at the so-called ``magic angle'' of $1.08^{\circ}$ \cite{cao2018unconventional,cao2018correlated}, and later in multilayered graphene stacks \cite{chen2019signatures,park2021tunable,park2022robust}, has attracted much attention.
Despite many interesting theoretical ideas to explain such phenomena, starting with the paradigm-setting model of Bistritzer and MacDonald (BM)~\cite{Bistritzer2011}, a simple real-space picture of the behavior of the low-energy electronic states remains elusive (for recent reviews, see Refs.~\cite{carr2020electronic,Aggarwal_2023}).  
Such a picture would be of great usefulness for building physically plausible theories of many-body effects, including quantum Hall effect states \cite{nuckolls2020strongly,wu2021chern}, superconductivity \cite{cao2018unconventional,lu2019superconductors} and the recent observations of states with fractional charge \cite{xie2021fractional}.

The electronic bands in tBLG are typically described using theoretical continuum models like the BM model and the chiral model \cite{tarnopolsky2019origin}, a special case of the BM model in which the hamiltonian has chiral symmetry. 
Such models capture the key features of tBLG, but only in an idealized, limiting-case sense. 
For instance, both models produce low-energy bands that are flat throughout the entire Brillouin Zone (BZ), perfectly flat in the case of the chiral model, and with vanishing Fermi velocity at magic angles, which in the BM model are $\theta = 1.05^{\circ}, 0.5^{\circ}, 0.35^{\circ}, 0.24^{\circ}, 0.2^{\circ}, \dots$.
These features have been assumed as essential elements in many other theoretical models which aim to explain the physics of tBLG.

More realistic descriptions of the electronic bands in tBLG have also been developed, which take into consideration the atomic relaxation in tBLG moir\'e structures. 
These descriptions include explicit first-principles calculations \cite{correa2014optical,uchida2014atomic,lucignano2019crucial} and \textit{ab initio} tight-binding (TB) models \cite{jung2014accurate,fang2016electronic,carr2018pressure}, based on TB hamiltonians with spatially modulated hoppings, as well as ``exact'' $\kp$ hamiltonians \cite{carr2019exact,fang2019angle,guinea2019continuum,kang2023pseudomagnetic,vafek2023continuum,miao2023truncated}, which perfectly reproduce the results from \textit{ab initio} TB models. 
In contrast to the idealized models, the more realistic first-principles models reveal that the bands are never perfectly flat or particle-hole symmetric, and that the Fermi velocity never actually reaches zero \cite{carr2019exact}. 
In addition, when lattice relaxation is taken into consideration, all of the magic angles are removed, except for the first one at approximately $1^{\circ}$, close to the experimentally observed range of values where superconductivity and correlated-insulator behavior have been reported \cite{cao2018unconventional,cao2018correlated}. 
The first-principles-based calculations also suggest that this magic angle is not a unique value, but rather a range of values in which the low-energy bands show optimal behavior in three respects: the Fermi velocity is minimized, the band width is minimized, and the band gaps separating the low-energy bands from the  valence and conduction manifolds are maximized. 

One route to a better understanding of exotic physics in tBLG is the development of models with the smallest possible number of bands which capture the \textit{realistic} features of the low-energy bands.
Several minimal phenomenological models have been developed to capture the low-energy physics of tBLG at small twist angles \cite{koshino2018maximally,kang2018symmetry,po2019faithful,carr2019exact,carr2019derivation,bernevig2021-I,song2022magic}.
Some of these are based on a minimal number of bands derived from effective Wannier orbitals on the moir\'e scale \cite{carr2019exact,carr2019derivation} and give reasonable agreement with band structures obtained from TB models and $\kp$ hamiltonians, using a small number of free parameters. 
However, there is an intrinsic arbitrariness associated with using such models to describe the low-energy bands in tBLG: the gauge freedom of the localized Wannier states can be used to tune the model so that any combination of these states can represent the low-energy bands. 
Because of this, it is impossible to determine the character of the low-energy bands using these phenomenological models. 
Moreover, the number of bands in these models (5, 8 or 10 \cite{po2019faithful,carr2019derivation}) is still prohibitively expensive for applications beyond single particle physics. 
Requiring the Wannier projection to produce the absolutely minimal set, namely a four-band model, including a pair of low-energy bands for each K-valley, results in so-called ``fidget spinner'' states \cite{koshino2018maximally,kang2018symmetry,po2018origin}, which are delocalized over several moir\'e cells.

Here, we revisit the properties of tBLG from a first-principles perspective. 
First, by examining the electronic bands and Fermi surface at half-filling over a very fine sampling of twist angles near the magic range, we identify several important features which can be used to characterize the magic range and which paint a richer picture of the physics in this range of twist angles. 
Second, we propose the simplest possible {\em real space} TB hamiltonian which captures these important features. 
Specifically, we propose a ``$(2+2)$''-band model, with two active low-energy bands and two auxiliary bands, whose origins are justified on the basis of the main structural features of the moir\'e supercell, as derived from atomistic relaxation calculations.
Finally, by projecting out the auxiliary bands, we derive an effective hamiltonian for the low-energy bands alone which, without sacrificing any accuracy in their description, affords a connection to a Hubbard-like model and describes their occupancy at half-filling. 

The rest of the paper is organized as follows: in Section \ref{sec:k-p} we provide a detailed study of the electronic states of tBLG in the magic range using the exact {\em ab initio} TB model developed in Refs.~\cite{carr2019exact} and \cite{fang2019angle}.
In Section \ref{sec:model} we describe the construction of the ``$(2+2)$''-band TB model and the effective hamiltonian of the minimal two-band model, as well as its connection to a Hubbard-like model. 
Finally, in Section \ref{sec:conclusion} we conclude with remarks on the relation of our minimal model to other theoretical descriptions of the low-energy bands.

\section{Origin and nature of low-energy bands
\label{sec:k-p}
}

\subsection{Effects of atomic relaxation}

The derivation of a minimal model has proven to be a non-trivial task because of the complexity of the underlying atomic structure in twisted bilayers with large moir\'e periods, that is, at twist angles of less than a few degrees; the actual atomic structure of tBLG for small angles does not consist of simply superimposing two pristine graphene lattices at the equilibrium interlayer separation, but includes significant atomic relaxation driven by energy minimization \cite{nam2017lattice,carr2018relaxation}.
Briefly, this is accomplished by defining the in-plane displacement vectors ${\bf U}^{(l)}({\rv})$ and the out-of-plane corrugation ${\bf h}^{(l)}({\rv}) = h^{(l)}(\rv) {\hat z}$ at an unrelaxed position $\rv$ of the supercell, where the index $l=t,b$ is the layer index (top and bottom).
The total energy is given by
\beq{}
   E^{\rm tot}[{\bf U}] =
   E^{\rm intra}[{\bf U}] + 
   E^{\rm inter}[{\bf U}]
\eec
expressed in terms of the displacement vector, where $E^{\rm intra}$ is the intralayer (in-plane) contribution, and $E^{\rm inter}$ is the interlayer (out-of-plane) contribution.
The first term is obtained from continuum elasticity theory in the linear approximation:
\beq{}
\begin{split}
E^{\rm intra}[{\bf U}] & = 
\sum_{l=t,b}
\int \dd^2 \rv \bigg\{
\frac{G}{2} \lb \nabla\cdot\bvec{U}^{(l)}\rb^2 \\
& + \frac{K}{2} \left[\lb \nabla^T\times\bvec{U}^{(l)}\rb^2 + \lb \nabla^T\cdot\bvec{U}^{(l)} \rb^2\right]\bigg\}
\end{split}
\eec
where $\nabla = (\partial_x,\partial_y)$, $\nabla^T = (\partial_y,\partial_x)$, and $G$ and $K$ are the shear and bulk modulus of monolayer graphene, respectively (obtained from first-principles calculations as $G = 9.0$ eV / \AA$^2$, and $K = 13.2$ eV / \AA$^2$).
The second term is obtained by employing the generalized stacking fault energy (GSFE) concept \cite{kaxiras1993free}, and is expressed as
\beq{}
E^{\rm inter}[{\bf U}] = 
\int \dd^2 \rv 
V^{\rm GSFE}({\bf s} (\rv) + {\bf U}^{(t)}(\rv) - {\bf U}^{(b)}(\rv) )
\eec
where ${\bf s}(\rv)$ is the local stacking at atomic position $\rv$, defined as the distance from an atom at $\rv$ in one layer to the position of the nearest neighbor of the same sublattice in the other layer \cite{carr2018relaxation}.
$V^{\rm GSFE}(\rv)$ is obtained by considering all possible relative displacements of two layers that span the entire {\em primitive} graphene unit cell and includes optimization with respect to interlayer separation $h(\rv) = |h^{(t)}(\rv) - h^{(b)}(\rv)|$. 
The values of $V^{\rm GSFE}(\rv)$ are obtained for a dense grid in real space and transformed by a Fourier expansion so that the final expression encompasses all the symmetries of the moir\'e supercell (for additional details, see Ref.~\cite{fang2019angle}). 
The relaxation is then obtained by minimizing the total energy with respect to the displacement fields ${\bf U}^{(l)}$, taking into consideration the symmetries of the system.

Atomic relaxation is known to play a role in many structural and electronic properties of multilayered materials~\cite{Falko_2011}. 
For instance, an important consequence of atomic relaxation in tBLG is that, although several magic angles are predicted by the BM model in the absence of atomic relaxation, only one magic angle has been observed experimentally. 
When lattice relaxation is taken into account, the other theoretically predicted magic angles vanish, leaving only one near the experimentally observed value \cite{carr2019exact}.

\begin{figure}[t!]
\centering
\includegraphics[width=\columnwidth]{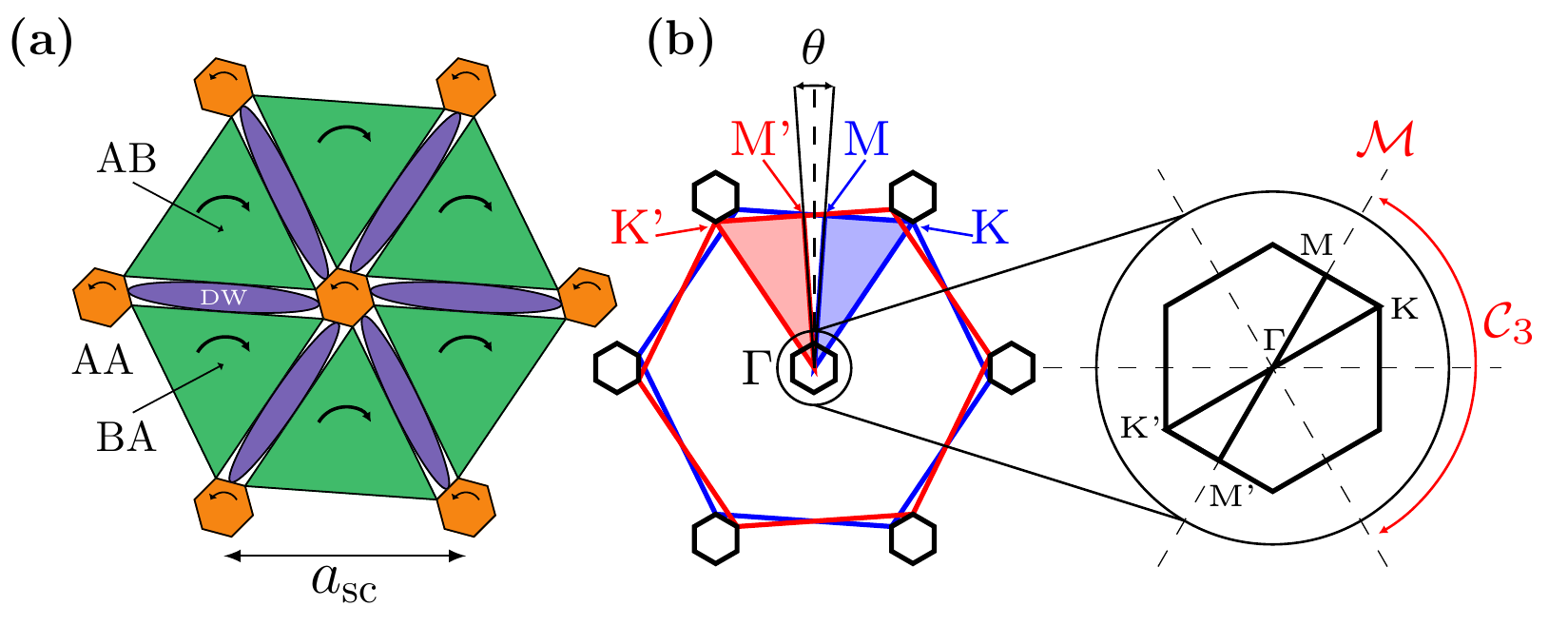}
\caption{
\textbf{(a)} Sketch of the local twisting and untwisting which occurs around the AA and AB/BA domain centers, respectively. 
The orange hexagons correspond to the AA regions, the green triangles to the AB/BA regions and the purple ellipses correspond to the DWs. 
$\aM$ is the moir\'{e} period.
\textbf{(b)} The monolayer BZs, in red and blue, which are rotated with respect to one another by $\th$. 
The high symmetry points $\G$, K/K' and M/M' are shown. 
The moir\'e BZ is the hexagon whose edges are formed by joining the K and K' points of the monolayers.
The magnified diagram shows the $\kv$-point paths considered and their $\C_3$ rotation symmetries and mirror planes $\M$.
}
\label{fig:relaxation}
\end{figure}

For twist angles $\lesssim 2^{\circ}$, atomic relaxation in tBLG results in the formation of three types of clearly identifiable domains, labeled AA, AB/BA and DW, see Fig.~\ref{fig:relaxation} (a).
In the energetically unfavorable AA domain, the atoms in the same sublattices are vertically aligned. 
In the energetically favored AB/BA domains (which are equivalent to each other by translation or rotation), half of the atoms in opposite sublattices are vertically aligned, and the other half are aligned with the the centers of the hexagonal rings of carbon atoms in the other layer.
In the domain wall stacking (DW), which separates the AB/BA stackings, the atoms in the two layers are offset by half a diagonal of the graphene primitive unit cell. 
The definitions of these three domain types correspond to infinite-size perfectly ordered regions, but in the moir\'{e} superlattice the three types of domains are connected, so the atomic alignments are close to those of the infinite regions in the majority of each domain and transition smoothly from one type to the other at the domain boundaries. 

As a result of relaxation, the AA regions tend to shrink to reduce their energy cost, while the AB/BA regions tend to increase in size to benefit the energy balance.  
The physics of this intricate atomic-scale reconstruction, which is responsible for the domains at the moir\'{e} scale, has been described using continuum elasticity by Zhang and Tadmor \cite{zhang2018structural} and used to explain experimental measurements of the patterns revealed by scattering \cite{yoo2019atomic}. 
Similar effects have been observed in other systems of twisted or strained bilayers and mulitlayers \cite{rakib2022moire, Falko_2011}.

For twist angles smaller than a critical value of $\th_{\rm c} \approx 1.2^{\circ}$, the sizes of the AA and DW domains reach a plateau, and only the AB/BA domains grow larger as the twist angle decreases.
Below this critical angle the moir\'e supercell can be represented as a combination of three intersecting lattices: a triangular lattice with sites at the centers of the small AA domains, an hexagonal (honeycomb) lattice with sites at the centers of the triangular AB/BA domains, and a Kagome lattice with sites at the centers of the the DW regions.

For small angles, in addition to changing the domain size, the local relaxation and strain relief results in an additional relative twist in the AA regions, $\Delta \th^{\rm AA}$ \cite{carr2018relaxation}. 
This results in a net local twist $\th_0^{\rm AA}$ that is {\em independent} of the global twist $\th$ imposed on the bilayer. 
At the same time, the AB/BA regions untwist by $\D\theta^{\rm AB/BA}$, i.e.~in the {\em opposite} sense from the global twist.
These relaxations are shown schematically in Fig.~\ref{fig:relaxation} (a).
Detailed atomistic-scale calculations predict a value of $\th^{\rm AA}_0 \approx 1.9^{\circ}$ \cite{zhang2018structural}.
From simple geometric considerations, it is straightforward to show that
\beq{eq:twist_angle}
\Delta \th^{\rm AB/BA} 
= \frac{\sqrt{3} \rho_{\rm AA}}{\sqrt{3} \rho_{\rm AA} - \aM} 
\D\theta^{\rm AA}
\approx
\frac{-\sqrt{3} \rho_{\rm AA}\D\theta^{\rm AA}}{a_0}  \theta
\eec
where $\rho_{\rm AA}$ is the radius of the AA region and $\aM$ is the moir\'{e} period; the last expression is valid in the limit $\rho_{\rm AA} \ll \aM$, and includes the relation $\aM = a_0/\theta$, where $a_0$ is the lattice constant of the primitive graphene unit cell. 
Using the values $\rho_{\rm AA}\approx 23$ \AA{} and 
$\D\theta^{\rm AA} \approx \theta_0^{\rm AA}\approx 1.9^\circ$ from Ref.~\cite{zhang2018structural}, for a global twist of $\theta =1.1^{\circ}$ this simple formula predicts $\D\th^{\rm AB/BA} = -0.35^{\circ}$, in close agreement with the multiscale simulations~\cite{zhang2018structural}. 
More generally, Eq.~(\ref{eq:twist_angle}) shows that the AB/BA regions are significantly {\em untwisted} for any twist angle ${\th \leq \th_{\rm c}}\approx 1.2^{\circ}$ due to relaxation, and as the last approximate expression shows the untwisting is proportional to $\theta$ with a constant of proportionality $\sim 0.53$.

We emphasize that the atomic relaxation is crucial in producing meaningful definitions for the AA, AB/BA and DW domains: without the relaxation, such domains would not exist, as their extent would be confined to an area of order a single unit cell of the bilayer graphene, and even then only approximately, while the transition from one type of region to another would be smooth and continuous over a length scale comparable to the moir\'e scale. 
This is actually the situation for larger ($\gtrsim 1.2^{\circ}$) twist angles, where atomic relaxation is negligible: moir\'e-scale domains do not exist, and as a consequence there is no interesting behavior, specifically no low-energy bands separated from the rest of the spectrum by band gaps.

Additionally, further bending/rippling of the domain walls has been proposed, which may lead to interesting effects such as the doubling of the moir\'e cell and the opening of mini-gaps \cite{rakib2022corrugation,rakib2023helical,kaliteevsky2023twirling}.
We also note that atomic relaxation is not confined to tBLG but is important in many bilayer systems. 
For instance, it plays a significant role in the vibrational properties of twisted bilayers \cite{koshino2019moire,ochoa2019moire,lu2022low,ochoa2022degradation}, and it underlies the appearance of ferroelectricity in bilayers that have a broken AB sublattice symmetry, such as hexagonal boron nitride (hBN) and transition metal dichalcogenides (TMDs)~\cite{yasuda2021stacking,zheng2020unconventional,bennett2022electrically,bennett2022theory}. 
In addition, atomic relaxation results in nontrivial topology, such as topologically protected 1D conduction channels along the DWs upon gating \cite{yoo2019atomic}, as well as real space topology from polarization \cite{bennett2023polar,bennett2023theory} and strain fields~\cite{engelke2023non}.

\subsection{Nature of low-energy bands}

\begin{figure*}[t!]
\centering
\includegraphics[width=\linewidth]{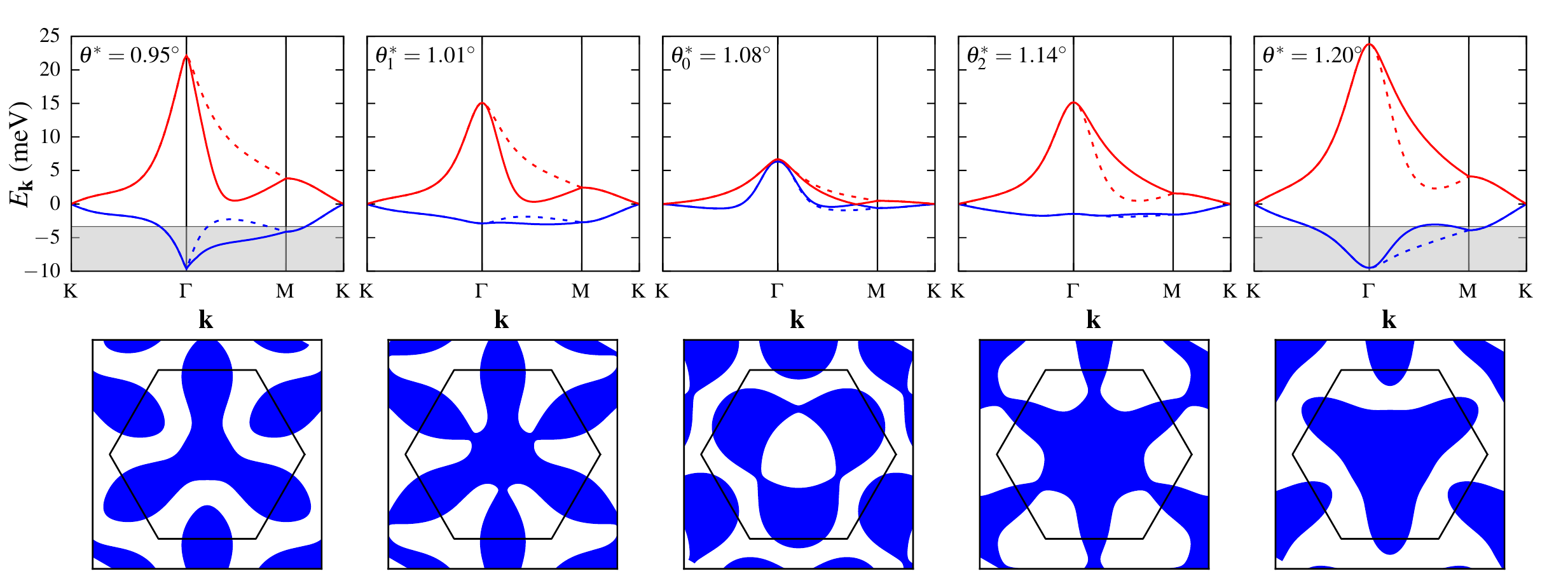}
\caption{
{\bf (Top)} Low-energy bands of tBLG along the K--$\G$--M--K (solid) and K'--$\G$--M'--K' (dashed) paths, for a selection of twist angles near the magic angle. 
The special angles $\th^*_1$ and $\th^*_2$, which define the magic range, as well as the angle $\th^*_0$, at which the bands are degenerate at $\G$, are shown. 
For the smallest and largest twist angles, the plots are shaded up to the half-filling lines of the lower bands.
{\bf (Bottom)} Fermi surfaces of the lower bands directly above, at half-filling.
}
\label{fig:kp-bands-fermi}
\end{figure*}

To examine the nature of the low-energy bands we rely on the $\kp$ model developed in Refs.~\cite{carr2019exact,fang2019angle}, which is a more realistic generalization of the BM model and similar models \cite{dos2007graphene,Bistritzer2011,mele2011band}. 
It is comprised of a pair of $2\times 2$ Dirac hamiltonians for the individual graphene layers as well as spatially modulated interlayer interactions due to the change in stacking configuration. 
The hamiltonian is written as a plane wave expansion about one of the K/K' valleys, which are related by exchanging the layers. 
Additionally, atomic relaxation effects are taken into account, and are described by a pseudo-gauge field \cite{manes2013generalized}, 
which is an in-plane correction, as well as momentum-dependent interlayer scattering terms, which are needed to capture the particle-hole asymmetry; this is the first-order correction to the interlayer hoppings being nonlocal \cite{fang2019angle}.
Although cast in a ${\kp}$ form, this model reproduces with excellent numerical accuracy the results of a full TB calculation with all atomic degrees of freedom, using the parameters that fit {\em ab initio} results, as derived in Refs.~\cite{carr2019exact,fang2019angle}.
Thus, in the following we refer to this model as the ``exact'' ${\kp}$ model to distinguish it from similar models based on heuristic arguments. 

Using the exact ${\kp}$ model, we obtain the low-energy bands of tBLG for a fine sampling of the twist angle $\theta$ close to the first magic angle of the BM model ($1.05^{\circ}$), along the two different paths in the moir\'{e} BZ shown in Fig.~\ref{fig:relaxation} (b).
We note that the model, being parameterized by first-principles density functional theory (DFT)
calculations, predicts the magic angle to be slightly lower than the experimentally observed one, although this does not affect the qualitative behavior.
For a more realistic comparison to experimental results, we compensate for
the limitations of DFT calculations 
by adjusting the twist angles by a constant shift of $\D\th_{\rm DFT} = 0.1^{\circ}$, that is, we 
report the results for the values 
of the twist angle 
$\th^* = \th + \D\th_{\rm DFT}$.
The low-energy bands are shown in Fig.~\ref{fig:kp-bands-fermi} for a few values of $\th^*$ that capture some salient features.
We first describe their behavior qualitatively. 
The bands resemble those of a honeycomb lattice, and are equal everywhere along the two paths except along $\G$--M/M', which is not enforced by any of the symmetries: $\C_3$, $\M$ or $\C_2\T$. 
As the twist angle decreases, the upper band becomes less dispersive, but never fully flat.
In comparison, the lower band becomes extremely flat and its curvature about $\G$ changes sign. 
The bands eventually touch at $\G$ and then open up again as the twist angle decreases further, with another sign change in the curvature of the lower band. 

\begin{figure}[t!]
\centering
\includegraphics[width=\columnwidth]{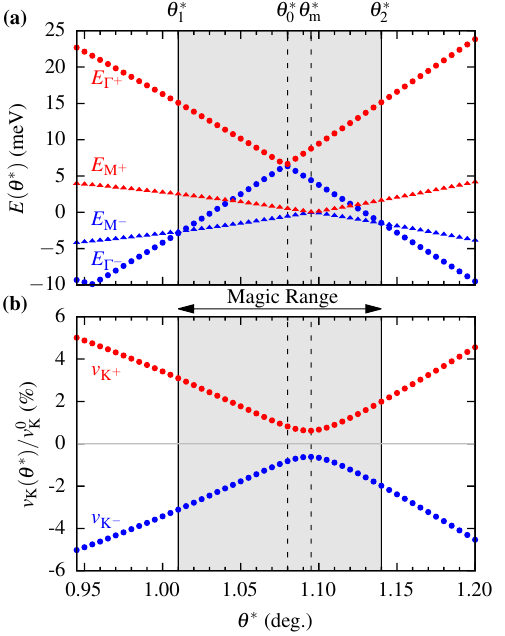}
\caption{
Magic range from $\kp$ bands.
{\bf (a)} Energy eigenvalues of the upper (red) and lower (blue) low-energy bands at the $\G$ and M points as a function of twist angle. The twist angles $\th^*_1$ and $\th^*_2$ at which the $\G$ and M eigenvalues of the lower band are equal are marked, and the region between them is shaded, indicating the magic range. 
The twist angles $\th^*_0$ and $\th^*_{\rm m}$, at which the eigenvalues of both bands at $\G$ and M are degenerate, respectively, are also shown.
{\bf (b)} Fermi velocity $v_{\rm K}(\th^*)$ of the upper (red) and lower (blue) low-energy bands as a function of twist angle, as a percentage of the Fermi velocity of graphene, $v^0_{\rm K} = 10^6$ m/s \cite{neto2009electronic}. 
The absolute value of the Fermi velocity of both bands reaches a minimum at $\th^*_{\rm m}$.
}
\label{fig:magic-range}
\end{figure}

For a more quantitative description of the low-energy bands, in Fig.~\ref{fig:magic-range} we show the behavior of the energy eigenvalues at high-symmetry points in the BZ, namely the values at $\Gamma$ and M, and the slope at K, i.e.~Fermi velocity $v_{\rm K}$.
The minimum bandwidth occurs in the lower band near two values of the twist angle where the eigenvalues at $\G$ and M become equal, namely $\th^*_{1}=1.01^{\circ}$, and $\th^*_{2}=1.14^{\circ}$.
We propose that the range between these angles be referred to as the ``magic range" because both bands remain quite flat (though still not \emph{completely} flat) throughout this region, as shown in Fig.~\ref{fig:magic-range} (a).
We note that the eigenvalues at $\Gamma$ and M never become degenerate for the top band, consistent with the fact that this band is always more dispersive (less flat) than the bottom band.

Within the magic range there are several features of interest.
The bands become degenerate at $\G$ for $\theta^*_0 = 1.08^{\circ}$, as shown in the middle panel of Fig.~\ref{fig:kp-bands-fermi} and by the crossing of the eigenvalues in Fig.~\ref{fig:magic-range} (a).
Additionally, the bands become degenerate at M for an angle $\theta^*_{\rm m}$ between 1.00$^\circ$ and 1.10$^\circ$, which is the same angle where the magnitude of the Fermi velocity reaches a minimum.
As described earlier, the Fermi velocity of both bands, which is positive for the top band and negative for the bottom band, never goes to zero, but its magnitude reaches a minimum at $\theta_{\rm m}$ and increases again away from this value.
The Fermi velocities of the top and bottom bands are equal in magnitude and opposite in sign throughout the region shown in Fig.~\ref{fig:magic-range}, including the entire magic range. 
Thus, a natural choice for the ``magic angle'' is to identify it with the value of $\theta^*_{\rm m}$ where the magnitude of the Fermi velocity of both bands acquires its minimum value and the eigenvalues at M become degenerate.  
Finally, we note that the magic angle is within the magic range, $\theta^*_{1} < \theta^*_{\rm m} < \theta^*_{2}$, although not exactly at its midpoint. 
Summarizing, the main features of the low-energy bands are:

\begin{itemize}
\item{there is always a pair (not counting K-valley and spin degeneracies) of low-energy bands near charge neutrality, which are degenerate at the K point of the BZ;}
\item{the two low-energy bands are always particle-hole {\em asymmetric}, with the lower band being generally flatter than the upper band;}
\item{the Fermi velocity at the K point of the BZ never goes to zero, but reaches a minimum at a special value denoted here by $\theta^*_{\rm m}$;}
\item{the top and bottom bands become degenerate at the M point of the BZ at $\theta^*_{\rm m}$;}
\item{neither band ever becomes exactly flat, but rather the curvature of the bottom band at $\G$ changes sign and the two bands become degenerate at $\Gamma$ at an angle $\theta^*_0$ close to $\theta^*_{\rm m}$;}
\item{the bottom band has the lowest dispersion (is most ``flat'') when the eigenvalues at $\G$ and M are equal, which occurs at two values of $\theta^*$, defined here as $\theta^*_{1}, \theta^*_{2}$.}
\end{itemize}

We next turn to the qualitative behavior of the band structure as revealed by the Fermi surface at half-filling of the low-energy bands.
This is important because Mott-insulating behavior in tBLG is observed experimentally at half-filling of each of the low-energy bands, and superconductivity is observed for small doping away from half-filling. 
In Fig.~\ref{fig:kp-bands-fermi} we show the Fermi surfaces at half-filling of the lower band for a few values of twist angle $\theta^*$. 
For $\th^* = 1.2^{\circ}$ the lower band forms a threefold surface around $\G$, with pockets around the K and K' points. 
The trigonal warping which results in the splitting along the $\G$--M/M' paths in the band structure plots is clear. 
The Fermi surface of the top band resembles that of the lower band, but with the orientation rotated by $\pi$. 
As the twist angle decreases and the lower band becomes flatter, the threefold surface becomes connected between neighboring BZs, but there still exist pockets around the K and K' points. When the curvature of the bottom band at $\G$ changes sign, a pocket opens at $\Gamma$. 
The bands along $\G$--M and $\G$--M' pass through one another, and the band along $\G$--M' falls below the half-filling line, causing the surface to rejoin, but with an orientation rotated by $\pi$. 
The orientation of the Fermi surface of the top band also has its orientation reversed as it passes through the magic range, indicating that the character of the bands has been exchanged.

\section{Model construction
\label{sec:model}
}

\begin{figure}[t!]
\centering
\includegraphics[width=\columnwidth]{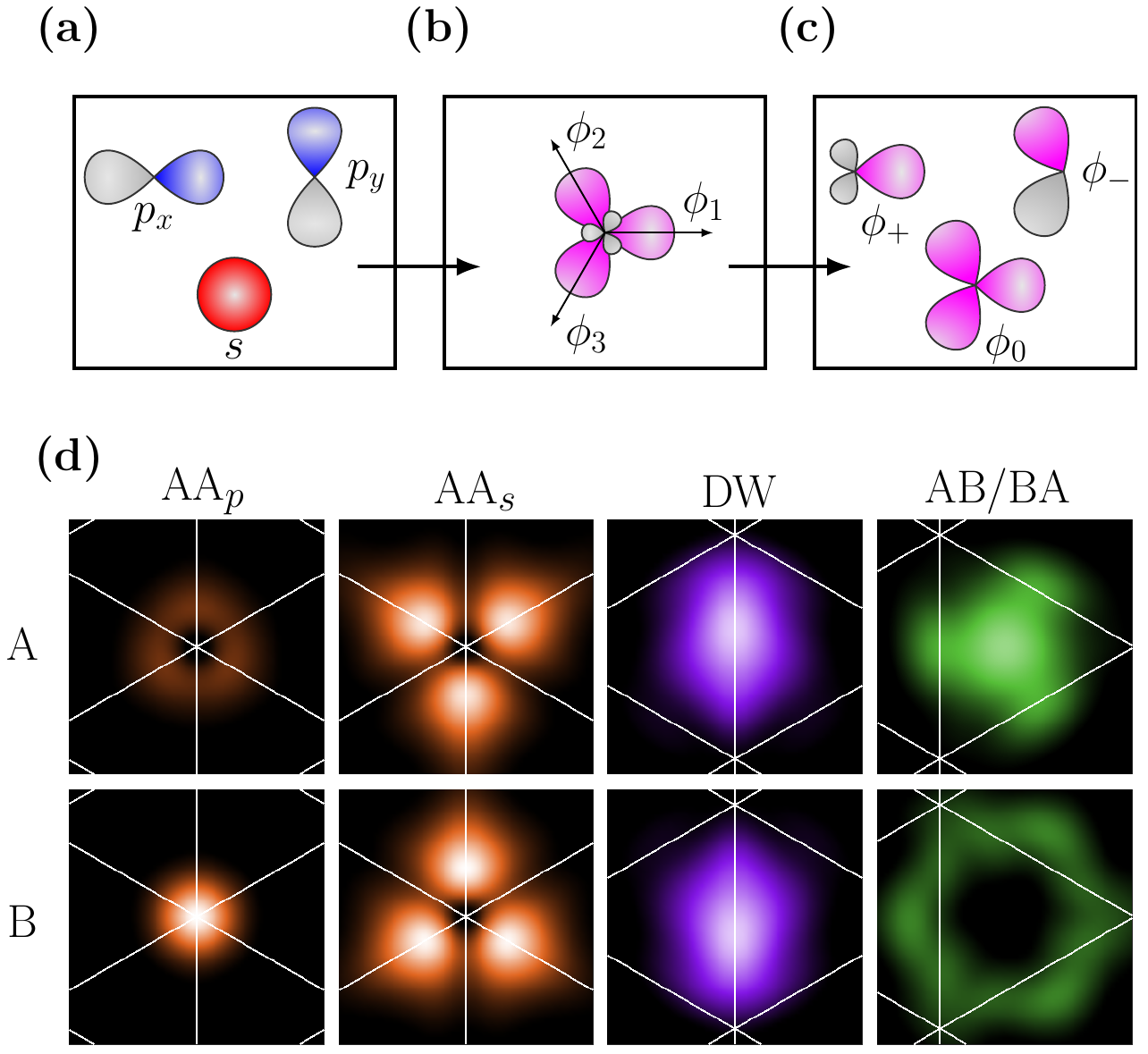}
\caption{
{\bf (a)} Illustration of the $s$ (red) and $p_x, p_y$ (blue) orbitals, and {\bf (b)} their linear combinations into the three $sp^2$ hybrid orbitals, labeled $\p_1, \p_2, \p_3$.
{\bf (c)} The $\p_0$, $\p_+$, $\p_-$ effective orbitals are shown on the right.
For each orbital, the lobes of positive sign are shown in magenta, and the lobes of negative sign are shown as gray.
{\bf (d)} Wavefunction magnitudes of the AA, DW and AB/BA states, projected onto the A and B sublattices from the ten-band model in Ref.~\cite{carr2019exact}, for $\th = 0.9^{\circ}$.
}
\label{fig:orbitals}
\end{figure}

\begin{figure*}[t!]
\centering
\includegraphics[width=1\linewidth]{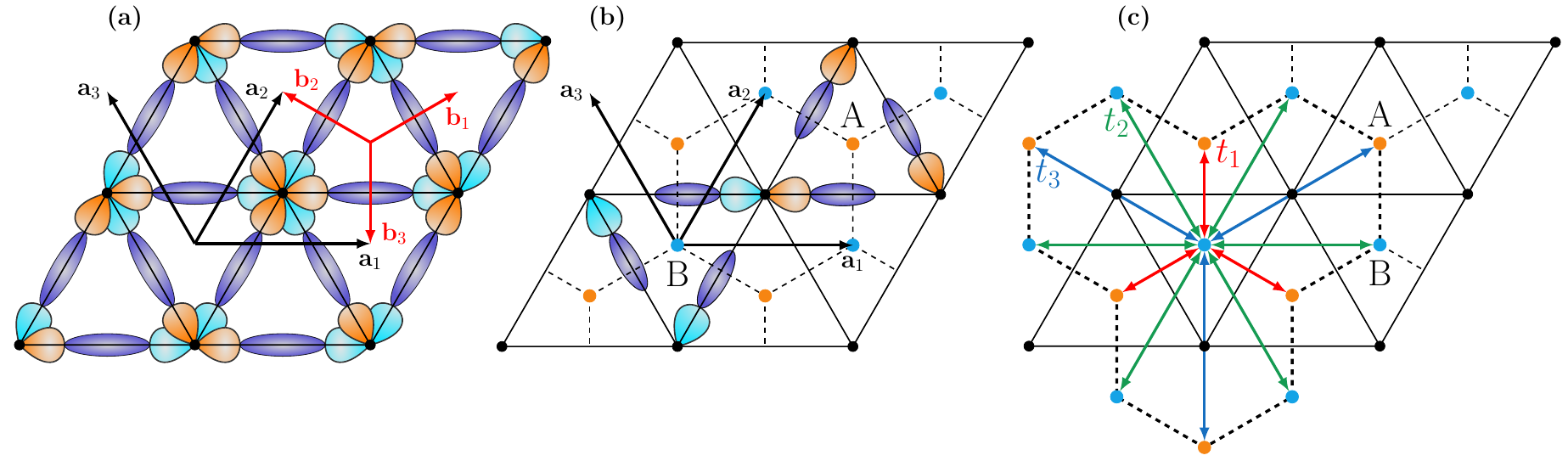}
\caption{
\textbf{(a)} Triangular lattice with the lattice vectors $\av_1$, $\av_2$ and $\av_3$ (black arrows), and the vectors $\bv_1$, $\bv_2$, $\bv_3$ (red arrows).  
A total of four unit cells of the periodic lattice are shown.
The central lattice site is decorated with one $\p_0$ orbital consisting of the three positive lobes of $\p_1, \p_2, \p_3$ for the A sublattice (orange), and a second orbital with the lobes rotated by $\pi$ for the B sublattice (light blue). 
The purple ellipses represent the DW states.
\textbf{(b)} Re-arrangement of the orbitals so that the lobes $\p_1, \p_2, \p_3$ are located at different lattice sites. 
The centers of the orbitals attributed to the A and B sublattices are shown, which form a honeycomb lattice, as indicated by the dashed black lines.
\textbf{(c)} Nearest neighbor hoppings included in the $(2+2)$-band hamiltonian, Eq.~\eqref{eq:H-4band}. 
The red arrows indicate the first nearest neighbor hoppings, between opposite sublattices, connected by vectors $\bv_i$.
The green arrows indicate the second nearest neighbor hoppings, on the same sublattice, connected by vectors $\av_i$.
The blue arrows indicate the third nearest neighbor hoppings, between opposite sublattices, connected by vectors $-2\bv_i$.
}
\label{fig:lattices}
\end{figure*}

\subsection{Real-space basis of effective orbitals}

We define the lattice vectors ${\av_1 = \aM\hat{x}}$ and ${\av_2  = \frac{\aM}{2}(\hat{x} + \sqrt{3}\hat{y})}$ which describe the moir\'{e} pattern, where $\aM$ is the moir\'e period, and $\av_3=\av_2-\av_1$, which is not a linearly independent vector but is introduced for convenience.
We also define the vectors ${\bv_1 = \frac{1}{3}(\av_1+\av_2)}$, ${\bv_2 = \frac{1}{3}(\av_2-2\av_1)}$, ${\bv_3 = \frac{1}{3}(\av_1-2\av_2)}$, which connect the centers of adjacent equilateral triangles in the triangular lattice or, equivalently, the A and B sublattice sites of the honeycomb lattice; these vectors are helpful in expressing certain terms in the TB hamiltonian.

Now we construct effective orbitals on the moir\'e scale, observing the following: the usual $sp^2$ hybrid orbitals, which are responsible for the in-plane $\s$ bonds in graphene, are obtained as linear combinations of the conventional atomic $s$, $p_x,p_y$ orbitals \cite{ek_jdj_2019}, see Fig.~\ref{fig:orbitals} (a).
These hybrid orbitals have a pronounced lobe of positive sign pointing in each of the three directions related by $\C_3$ rotations, and a smaller lobe of negative sign pointing in the opposite direction, see Fig.~\ref{fig:orbitals} (b).

We can consider more general single-lobe hybrid orbitals as the basis which will generate effective states $\p_0$, $\p_{+}$ and $\p_{-}$, of $s$, $p_{x}$ and $p_{y}$ character, but which maintain the directional features, as shown schematically in Fig.~\ref{fig:orbitals} (c).
In particular, we will work with two such orbitals of $s$ character, namely $\p_0^{\rm A}, \p_0^{\rm B}$, which possess prominent directional features: the first one is as shown in Fig.~\ref{fig:orbitals} (c) and the second is rotated by $\pi$, both centered at the AA sites of the moir\'e lattice.
Considering both states, the wavefunction has $f$-like character at the AA sites, echoing the model in Ref.~\cite{song2022magic}, where the states attributed to the low-energy bands were compared to the $f$ electrons in heavy fermion superconductors.
This is a good description of the character of the AA sites, as shown from phenomenological models with a small number of effective orbitals \cite{carr2019exact}, see Fig.~\ref{fig:orbitals} (d).
The DW states form a Kagome lattice of elongated ellipses, and the AB/BA domains form a honeycomb lattice with wide triangles of opposite orientation. 
The AA and DW states, which describe the low energy bands \cite{carr2019exact}, are illustrated in Fig.~\ref{fig:lattices}(a).

The AA states can also be expressed as:
\beq{eq:moire-hybrids}
\p_0^{\rm X}(\rv) = 
\frac{1}{\sqrt{3}} 
\sum_{i=1}^3 
\p_i^{\rm X}
(\rv \pm \bv_i - \rv^{\rm X}), 
\quad {\rm X}={\rm A,B}
\eec
where $\p_i^{\rm X}(\rv)$ are the single-lobe orbitals, in analogy to the $sp^2$ hybrids, with the upper and lower signs corresponding to the A-type and B-type lobes, respectively.
The interesting aspect of these states is that the three lobes of a single orbital are located at the three different corners of the triangle whose center is located at $\rv^{\rm X}$.
Choosing $\rv^{\rm A}=2\bv_1$ and $\rv^{\rm B}=0$ yields the arrangement of single lobes shown in Fig.~\ref{fig:lattices} (b), which, when repeated periodically on the Bravais lattice defined by the vectors $\av_1$, $\av_2$, produces a distribution of $\p_0$-type orbitals identical to that shown in Fig.~\ref{fig:lattices} (a). 
The states defined in Eq.~\eqref{eq:moire-hybrids} form a natural basis of a {\em honeycomb} lattice, whose dominant hopping matrix elements are to the Kagome states located along the DWs of the moir\'e pattern, that is, on the lines connecting the sites of the triangular lattice.

\subsection{$(2+2)$-band hamiltonian}

\begin{figure*}[t!]
\centering
\includegraphics[width=1\linewidth]{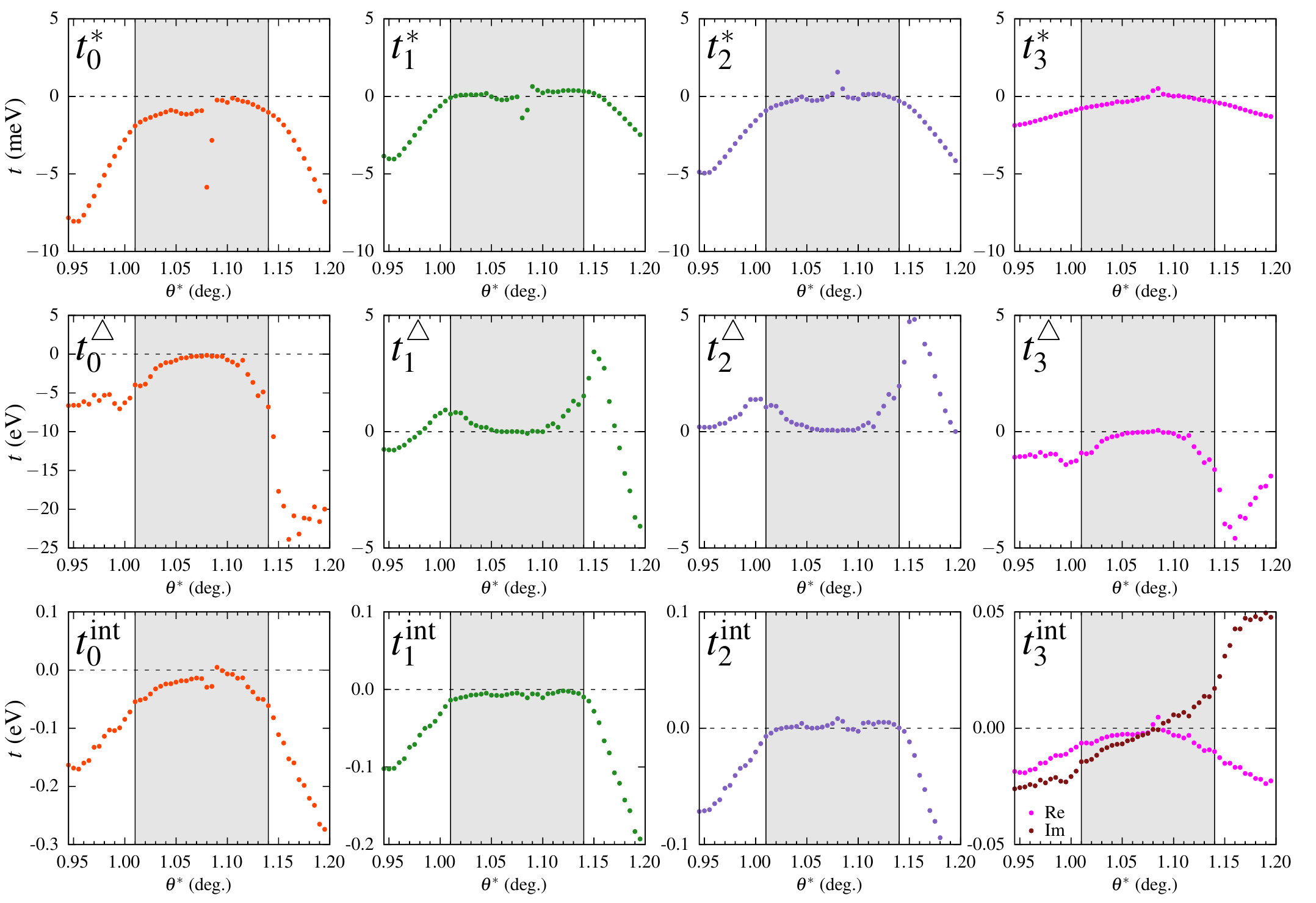}
\caption{
Values of the hopping parameters $t^{\ast}_i$, $t^{\triangle}_i$ and $t^{\rm int}_i$, $i=0,1,2,3$, that appear in $\H^{\ast}$, $\H^{\triangle}$, and $\H^{\rm int}$, respectively, as determined by fitting to the $\kp$ low-energy bands, for the range of twist angles $0.95^{\circ} \leq \theta^* \leq 1.2^{\circ}$.
The shaded regions indicate the magic range.
}
\label{fig:fitting}
\end{figure*}

Having argued that a natural set of orbitals on a honeycomb lattice are the $\p_0$ states defined in Eq.~\eqref{eq:moire-hybrids}, we propose a four-band model with the following structure:
\beq{eq:H-4band}
\H^{(2+2)}_{\kv} = 
\begin{pmatrix}
\H^{\ast}_{\kv} & 
\H^{\rm int}_{\kv} \\
 \H^{\rm int, \dagger}_{\kv}& \H^{\triangle}_{\kv}
\end{pmatrix}
\eep
This hamiltonian is comprised of a $2\times 2$ sub-matrix $\H^{\ast}_{\kv}$ of $\p_0$ states at the AA sites, described by a honeycomb lattice, a second $2\times 2$ sub-matrix $\H^{\triangle}_{\kv}$ for the symmetric combinations of the DW states which mediate the hoppings between the AA states, which also form a honeycomb lattice, and an off-diagonal $2\times 2$ interaction matrix $\H^{\rm int}_{\kv}$ that connects the two honeycomb lattices. 
The AA states describe the low-energy bands, which is natural as the DOS at charge neutrality is largest around the AA sites, while the DW states provide complementary (or auxiliary) bands, which are necessary to ensure the correct symmetries of tBLG \cite{ahn2019failure}. 
The first sub-matrix in $\H^{(2+2)}_{\kv}$ for the AA states is given by
\beq{eq:H-AAA}
\H^{\ast}_{\kv} = 
\begin{pmatrix}
t^{\ast}_0 + t^{\ast}_2 f_2(\kv) &  t^{\ast}_1 f_1(\kv) + t^{\ast}_3 f_3(\kv) \\[5pt]
 t^{\ast}_1 f_1^{\dagger}(\kv) + t^{\ast}_3 f_3^{\dagger}(\kv) & t^{\ast}_0 + t^{\ast}_2 f_2(\kv)
\end{pmatrix} 
\eec
where the diagonal elements represent the hoppings within each sublattice, up to second nearest neighbors of the honeycomb lattice, and the off-diagonal terms represent the interactions between sublattices, up to third nearest neighbors, as shown in Fig.~\ref{fig:lattices} (c). 
The scalar functions $f_i(\kv)$ are given by
\beq{eq:f-functions}
\begin{split}
f_1(\kv) & = \sum_{j=1}^{3} \exp\lb i \kv\cdot\bv_j 
\rb,  \quad 
f_2(\kv) = \sum_{j=1}^{3} \cos\lb \kv\cdot\av_j \rb, \\
f_3(\kv) & = \sum_{j=1}^{3} \exp\lb -2i \kv\cdot\bv_j \rb, \\
\end{split}
\eeq
which describe the interactions between first, second and third nearest neighbors on the honeycomb lattice. 
An arbitrary number of further neighbor interactions could be included to systematically fit to and reproduce the bands from the $\kp$ hamiltonian, but we find that the main features of the low-energy bands are reproduced by including up to third nearest neighbor interactions.
We take the hamiltonian of the DW states, $\H^{\triangle}_{\kv}$, to be of exactly the same form as the low-energy bands hamiltonian, $\H^{\ast}_{\kv}$, only with different parameters, namely $t^{\triangle}_i, i=0,1,2,3$.
The hamiltonian describing the interaction between the two lattices, $\H^{\rm int}_{\kv}$, also has the same form as $\H^{\ast}_{\kv}$ and $\H^{\triangle}_{\kv}$, with another set of parameters, $t^{\rm int}_i, i=0,1,2,3$, but differs from those two hamiltonians in that the parameter $t^{\rm int}_3$ is allowed to be complex. 
It turns out that a complex phase is required to reproduce the differences between the bands along the $\G$--M and $\G$--M' paths.

This model contains a total of 13 independent parameters, which can be fit to reproduce the low-energy bands of tBLG obtained from the exact $\kp$ model.
We used {\sc mathematica} to determine the hopping parameters by minimizing the sum of squares of the differences between the low-energy bands in the $\kp$ model and the $(2+2)$-band model for the range of twist angles shown in Fig.~\ref{fig:magic-range}.
The auxiliary bands, taken to be the valence bands directly below the low-energy bands for each twist angle, were not included in the fitting. 
The fits could be improved arbitrarily by including more nearest neighbor interactions, although this is not pursued here as the goal is to reproduce the main features of the low-energy bands using a minimal model.

The parameters obtained from fitting to the $\kp$ bands are shown in Fig.~\ref{fig:fitting} as a function of $\th^*$, and the resulting low-energy bands are provided in the Supplementary Material (SM).
Several conclusions can be drawn about the behavior of the parameters in the model and their physical meaning.
First, we note there are three separate energy scales in the three different $2\times 2$ block hamiltonians.
The hoppings in $\H^{\ast}$, which describe the low-energy bands, are of order 10 meV, and the hoppings in $\H^{\triangle}$, are of order 1 eV.
Qualitatively, the latter hoppings determine the position and dispersion of the auxiliary bands.
This does not influence the low-energy bands much outside of the magic range, but inside the magic range, the parameters in $\H_{\triangle}$ decrease to values of order 1-10 meV, and the distance between the low-energy bands and auxiliary bands is minimized, causing the curvature of the lower band to change.
The hoppings in $\H^{\rm int}$, describing the interactions between the two lattices, are of an energy scale between those of $\H^{\ast}$ and $\H^{\triangle}$, of order 10-100 meV.
We note that the imaginary part of $t^{\rm int}_3$ determines the splitting along the $\G$--M/M' paths and hence the orientation of the Fermi surfaces; there is a change in the sign of the imaginary part of $t^{\rm int}_3$ as $\theta^*$ spans the magic range. 
This sign change is responsible for the change in orientation of the Fermi surface shown in Fig.~\ref{fig:kp-bands-fermi}.

\subsection{Towards a Hubbard-like model}

\begin{figure*}[t!]
\centering
\includegraphics[width=\linewidth]{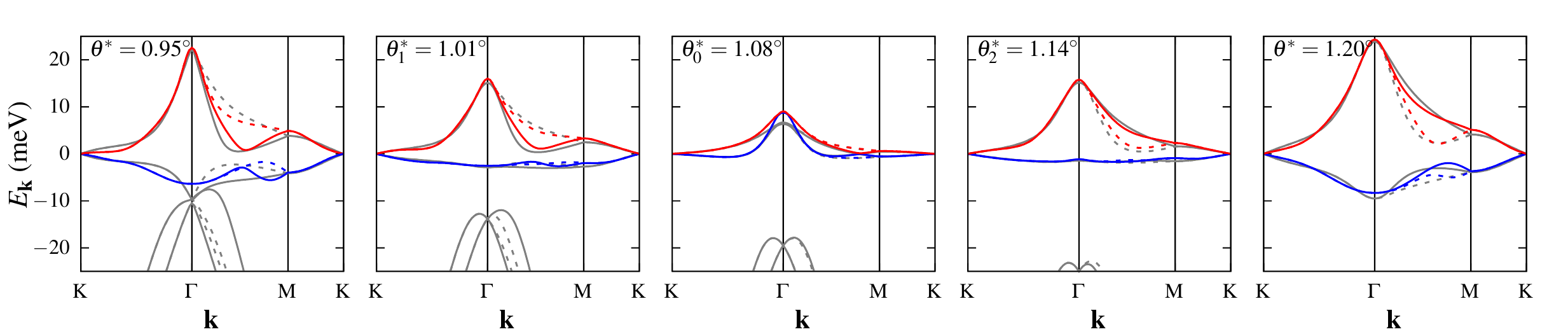}
\caption{
Low-energy bands obtained by solving Eq.~\eqref{eq:H-eff-a}, setting $E=0$ in Eq.~\eqref{eq:H-eff-b}, using the fits to the $(2+2)$-band hamiltonian in Fig.~\ref{fig:fitting}.
The $\kp$ bands are shown in gray.
The solid and dashed lines show the bands along the K--$\G$--M--K and K'--$\G$--M'--K' paths, respectively.
}
\label{fig:fit}
\end{figure*}

An eigenvector of the hamiltonian $\H^{(2+2)}_{\kv}$ can be represented by a four-component spinor, $\Phi$, which can be broken in the components $(\phi, \phi')$, each of them being a two-component spinor and representing predominantly the low-energy states $(\phi)$ and the highly dispersive states $(\phi')$, although the presence of the interaction term mixes the two sets:
\beq{eq:4by4_matrix}
\H^{(2+2)}_{\kv} \Phi = E\Phi
\implies
\begin{pmatrix}
\H^{\ast}_{\kv}
& \H^{\rm int}_{\kv} \\
 \H^{\rm int, \dagger}_{\kv}
 & \H^{\triangle}_{\kv}
\end{pmatrix}
\begin{pmatrix}
    \phi \\ \phi'
\end{pmatrix}
 = E 
\begin{pmatrix}
    \phi \\ \phi'
\end{pmatrix}
\eep
We may consider the dispersive states as a ``bath'' to which the low-energy states are coupled, in addition to the hopping matrix elements among themselves.
This picture can be taken one step further by eliminating the dispersive bands to obtain the {\em effective} $2\times 2$ hamiltonian that couples only the low-energy states. 
This effective hamiltonian is derived formally from the set of two equations in $\phi$,$\phi'$, implied by Eq.~\eqref{eq:4by4_matrix}.
By solving the second equation for $\phi'$ in terms of $\phi$ and substituting the resulting expression into the first equation we obtain the energy-dependent effective hamiltonian equation for $\phi$ in terms of $\H^{\ast}_{\kv}$, $\H^{\triangle}_{\kv}$ and $\H^{\rm int}_{\kv}$:
\beq{eq:H-eff-a}
\H^{\rm eff}_{\kv}(E) \phi 
= E\phi
\eec
where
\beq{eq:H-eff-b}
\H^{\rm eff}_{\kv}(E) 
= \H^{\ast}_{\kv} - 
\H^{\rm int}_{\kv} 
\left( \H^{\triangle}_{\kv} - E\right)^{-1} \H^{\rm int, \dagger}_{\kv}
\eep
Given that the scale of the parameters $t^{\triangle}_i$ is eV while the scale of the low-energy bands is meV (see Fig.~\ref{fig:fitting}), neglecting the energy dependence of $\H^{\rm eff}_{\kv}(E)$ may be a reasonable approximation, as it enters in the effective hamiltonian only through the combination $(\H^{\triangle}_{\kv}-E)$. 

The effective hamiltonian $\H^{\rm eff}_{\kv}(E)$, even in the approximation mentioned above where its energy dependence is neglected (equivalent to setting $E=0$ in Eq.~\eqref{eq:H-eff-b}), provides an excellent description of the low-energy bands, as shown in Fig.~\ref{fig:fit} and the SM.
However, one disadvantage is that the $\kv$-dependence of the effective hamiltonian is non-trivial, with the functions $f_i(\kv)$ of Eq.~\eqref{eq:f-functions} entering non-linearly in the second term of $\H^{\rm eff}_{\kv}$.
This is in contrast to the case of the full $(2+2)$-band hamiltonian, where the $\kv$-dependence of each term is clear and physically motivated. 
For some values of $\kv$ the problem simplifies, as shown in the SM.

For the creation of a faithful physical Hubbard-type model, one needs to specify the values of the parameters that enter in the effective hamiltonian defined in Eq.~\eqref{eq:H-eff-b}, as well as the on-site Coulomb repulsion terms
for the various sectors. 
These values will also determine the type of model appropriate for the system under consideration (for a recent review of Hubbard models, see Ref.~\cite{Arovas_2022}). 
As mentioned earlier, Eq.~\eqref{eq:H-eff-b} contains non-trivial $\kv$-dependence, while the estimation of various Coulomb repulsion terms ( $U$'s) 
requires careful treatment of the actual orbitals involved (see, for example, Ref.~\cite{koshino2018maximally}).

While our model reproduces the {\it ab initio} energy bands, it does not contain information about the actual wavefunctions in terms of atomic orbitals, which would be needed for reliable estimation of the $U$ terms and related properties.
Nevertheless, some properties may be inferred from the nature of the energy bands. 
Specifically, since our model is comprised of two honeycomb lattices, the geometric properties are expected to be similar to those of graphene: a Berry curvature which diverges at the K and K' points, and a Zak phase of $\pm \pi$ obtained upon integrating the connection around a contour enclosing the K/K' point (see the SM for the behavior of the bands near the K/K' points). Previous works have developed moir\'e-scale effective models (Refs.~\cite{po2019faithful,carr2019derivation}), but have not been used to study band topology. For this, we believe a more detailed and rigorous treatment would be needed, such as in the models in Refs.~\cite{kang2023pseudomagnetic,vafek2023continuum,po2019faithful,carr2019derivation,bernevig2021-I,song2022magic}, full atomistic tight-binding models, or large-scale first-principles calculations.

\section{Discussion and conclusions
\label{sec:conclusion}
}

In the present work, we revisited the problem of the low-energy bands in tBLG with an aim to better understand their origin and to make connections to correlated electron behavior in this system. 
We first reviewed the effects of atomic relaxation which is the driving force for creating different types of electronic states associated with the AA regions, the DW regions and the AB/BA regions of the moir\'e supercell.
We then examined in detail the behavior of the electronic low-energy bands of tBLG near the magic range, as obtained from {\em ab initio} TB methods. 
These bands exhibit more interesting features than the Fermi velocity simply going to zero at a magic angle, as predicted by idealized continuum models~\cite{Bistritzer2011}. 
In fact, the Fermi velocity never vanishes, but instead attains a finite minimum value~\cite{carr2019exact}.
The low-energy bands exhibit the smallest dispersion when the eigenvalues at the $\G$ and M points of the supercell BZ become equal.
We propose that the values at which this occurs serve as a proper definition for the bounds of the magic range of twist angles.
Within this range, the curvature of the lower band at $\G$ is \emph{negative} and it even intersects the upper band at $\G$ near the middle of the magic range. 
The two bands also touch at M at the same twist angle where the Fermi velocity reaches a minimum, near the center of the magic range.
The bands become more dispersive again at twist angles below the magic range, although the orientation of the bands has changed, which is evident from the shape of the Fermi surfaces at half filling.

Overall, our analysis of the low-energy band features suggests that more attention should be paid to their behavior in the neighborhood of the $\G$ and M points of the supercell BZ, and for a range of twist angles $\theta^*_{1,2} \approx \theta^*_{\rm m} \pm 0.07^{\circ}$, where $\theta^*_{\rm m}$ is the magic angle at which the Fermi velocity attains its (non-zero) minimum value; at the boundaries of the magic range the low-energy bands become most flat due to the degeneracy of the $\Gamma$ and M points, and the Fermi surface at half filling exhibits intriguing behavior.
In fact, we find that at half filling of the lower band the K/K' valleys remain {\em unoccupied} throughout the magic range of twist angles. Close to the middle of the magic range (where the Fermi velocity reaches a minimum), even the $\Gamma$ valley is unoccupied, and only states near the M point are involved. 

Additionally, we proposed a $(2+2)$-band model (per spin and layer) which captures the main features of the low-energy bands throughout the magic range. 
The model is comprised of two honeycomb lattices, one for the AA sites which describe the low-energy bands, and another for the DW states which serve as the auxiliary bands necessary to capture the symmetries of the low-energy bands~\cite{po2018origin,ahn2019failure}.
Our model is physically motivated by the form of the wavefunctions obtained from {\em ab initio}-based models, and contains a small number of physically intuitive parameters.
This model gives a satisfactory description of the low-energy bands of exact $\kp$ hamiltonians, and provides insight into the dramatic twist-angle-dependence of those bands, particularly within the magic range.

The motivation for choosing 4 bands in a minimal model is based on two facts: First, as is clear from the {\em ab initio} tight-binding bands discussed in Section II.B, the most interesting states for studying correlated electron behavior (the ``flatest'' bands) are the valence bands of the moir\'e supercell, which are in close proximity to the highly dispersive bands with lower energy, as seen clearly in Fig. \ref{fig:fit}; thus the 2 low-energy bands and 2 of the adjacent dispersive bands are all that is required to capture this behavior near the charge neutrality point (CNP). 
Second, as more detailed studies of the symmetries of the bands reveal ~\cite{po2019faithful,carr2019derivation}, it is possible to produce minimal models that contain as few as 5 bands by focusing on one sector only (conduction or valence bands, relative to the CNP) through Wannerization of the {\em ab initio} tight-binding results; our model represents an attempt to further reduce this number to 4, by changing the underlying lattices of both the AA-orbitals (usually taken to be a triangular lattice) and the DW-orbitals (usually taken to be a Kagome lattice), to a honeycomb lattice.
These two facts allow us to focus on the valence sector of the moir\'e bands and neglect the higher energy bands, without sacrificing any critical aspect of the behavior. 

In an attempt to provide a link to a Hubbard-type model that can capture the many-body aspects of the system, we projected the auxiliary bands out of the $(2+2)$-band hamiltonian to produce an effective $2\times 2$ hamiltonian for the low-energy bands. 
Although the $\kv$-dependence of the matrix elements of the effective hamiltonian is not as physically transparent as in the original $(2+2)$-band hamiltonian, the former still offers an excellent description of the two low-energy bands. 
Building an accurate Hubbard-type model from these states lies beyond the scope of the present paper and is left for future publications. 

Lastly, here we compare and contrast the essential features of our model to related earlier work. 
We first emphasize that our model applies to the behavior of the bands in the magic range of angles, which has a width of $0.14^{\circ}$, consistent with experimental results~\cite{cao2018unconventional}, 
where superconductivity was observed at twist angles of 1.05$^{\circ}$ and 1.16$^{\circ}$, a fact which is often overlooked. 
In contrast, other minimal models, for instance Refs.~\cite{po2019faithful,koshino2019moire}, constructed a model valid for a single magic-angle value based on two pairs of low-energy bands for the two different valleys.
Our model captures a wide range of behavior for a very fine sampling of twist angles within the magic range where the bands change rapidly.
The parameters that enter in our model have physically motivated meaning, and evolve smoothly as a function of twist angle (except at $\th^*_{0}$, where the bands touch at $\G$), which is quite remarkable for a simple $4\times 4$ hamiltonian.

Our model suggests that a natural description of the AA sites consists of two orbitals interacting on a honeycomb lattice rather than two orbitals on the triangular lattice formed by the AA sites. 
The orbitals of our model have lobes distributed over the corners of a triangle formed by AA regions, a feature similar to previously considered models of the low-energy states, see Refs.~\cite{po2018origin,koshino2018maximally}. 
However, an important difference with those previous works is that the distributed lobes of our model point {\em along the sides} of the triangle, whereas those of the previous models point {\em toward the center} of the triangle. 
This has the consequence that in our model the primary low-energy orbitals, consisting of the three distributed lobes, couple predominantly to the states of the Kagome lattice formed by the DW of the moir\'e pattern. 
The latter states represent the highly dispersive, auxiliary bands and can be described as consisting of elongated ellipses whose linear combination form orbitals also centered at the same honeycomb lattice as the primary three-lobe states. 
In this fashion, all the important states exist on a common honeycomb lattice.

The coupling of the low-energy bands to auxiliary dispersive bands has been proposed recently by Bernevig and co-workers, see Ref.~\cite{song2022magic}.  
The difference from those works is that in the model proposed here both the low-energy bands and the dispersive bands are physically motivated by features of the moir\'e pattern and the associated localized orbitals obtained from models with a much larger number of bands~\cite{carr2019exact,carr2019derivation} that accurately reproduce the {\it ab initio} bands. 
Interestingly, the low-energy orbitals we derive from these considerations have a shape consistent with the $f$-orbital shape associated with the AA sites in the model of Ref.~\cite{song2022magic}.

Another important difference between the models presented here and similar attempts to focus on the low-energy bands is that our bands are strongly electron-hole {\em asymmetric}.  
This is consistent with {\it ab initio} calculations~\cite{fang2016electronic} as well as with experiment~\cite{cao2018correlated}.
The particle-hole asymmetry of the bands has important consequences for their behavior. 
Specifically, the lower of the low-energy bands is much more flat than the upper one, again consistent with experimental indications that superconductivity is much more pronounced for hole doping than for electron doping~\cite{cao2018unconventional}.

Of course, some compromises need to be made in order to describe the low-energy bands with a few-band model. 
The low-energy states cannot be localized into fewer than 5 bands (per spin and valley)~\cite{carr2019derivation}, 
and this is clearly the case here, as both the lobes at AA sites
and the domain-wall states are distributed across a moir\'e cell. 
This is consistent with the observation that the bands near the magic angle are predicted to have fragile topology~\cite{po2019faithful,bernevig2021-I}. 
Additionally, the wavefunctions of the two-band model no longer satisfy all of the symmetries of tBLG; 
for instance, the lobes at AA sites, when distributed across a unit cell, no longer satisfy the $\M$ symmetry. 
If one insists on preserving these symmetries, then a larger number of bands is necessary which makes the formulation of a many-body model intractable.
Alternatively, it may be possible to 
restore some of the symmetries by considering a supercell of the moir\'e cell, as recent experimental evidence seems to suggest~\cite{Yazdani2023}.

In summary, the model of the low-energy bands derived here offers certain advantages in that it describes the states of interest accurately and with the use of relatively few and physically motivated parameters, whose values evolve smoothly as a function of the twist angle. 
This model may serve as a useful springboard for capturing the correlated electron behavior in tBLG. 

\section*{Acknowledgments}

We have benefited from helpful discussions with Ziyan Zhu 
and from a critical reading of the manuscript by Efstratios Manousakis.
We acknowledge support from a grant from the Simons Foundation award No.~896626, the US Army Research Office (ARO) MURI project under grant No.~W911NF-21-0147, and the National Science Foundation DMREF program under award No.~DMR-1922172.


%

\clearpage



\section*{Supplementary material}

## MODEL BANDS COMPARED TO THE $\mathbf{k} \cdot \mathbf{p}$ SPECTRUM

Fig. 1 shows the low-energy eigenvalues of the $(2+2)$-band hamiltonian, $\mathcal{H}_\mathbf{k}^{(2+2)}$, calculated using the parameters determined by fitting to the $\mathbf{k} \cdot \mathbf{p}$ spectrum, for twist angles $\theta^*$ between $0.95°$ and $1.20°$.

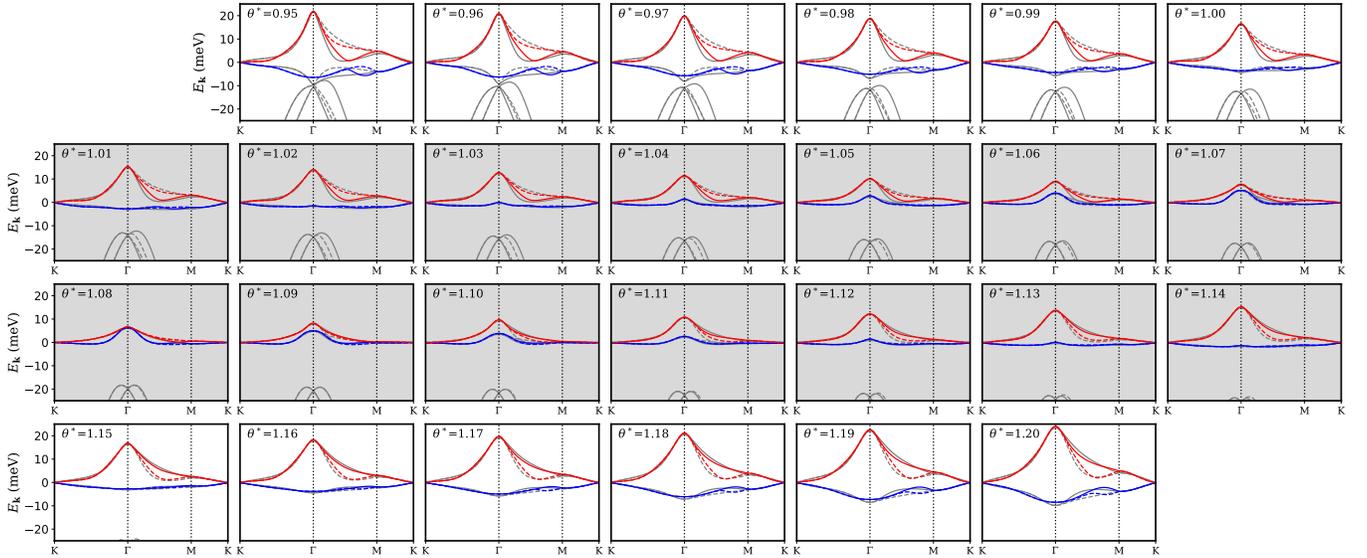

FIG. 1. Low-energy bands of the $(2+2)$-band hamiltonian for angles $\theta^*$ between $0.95°$ and $1.20°$ calculated using the fit parameters described in the main text (red and blue), compared to the bands from the $\mathbf{k} \cdot \mathbf{p}$ model (gray). A light gray background indicates angles that lie within the magic range.

Fig. 2 shows the energy eigenvalues of the two-band effective hamiltonian, $\mathcal{H}_\mathbf{k}^{\text{eff}}(E=0)$, for twist angles $\theta^*$ between $0.95°$ and $1.20°$, calculated using the the same fit parameters shown in the main text. The main differences from Fig. 1 are at the $\Gamma$-point in the magic range.

## ANALYTIC RESULTS NEAR THE K/K' POINTS

The $(2+2)$-band hamiltonian $\mathcal{H}^{(2+2)}$ can be expressed in terms of an energy-dependent effective hamiltonian as described in the main text:

$$\mathcal{H}_\mathbf{k}^{\text{eff}}(E) = \mathcal{H}_\mathbf{k}^* - \mathcal{H}_\mathbf{k}^{\text{int}} \left(\mathcal{H}_\mathbf{k}^{\triangle} - E\right)^{-1} \mathcal{H}_\mathbf{k}^{\text{int},\dagger}. \tag{1}$$

---

* kaxiras@physics.harvard.edu

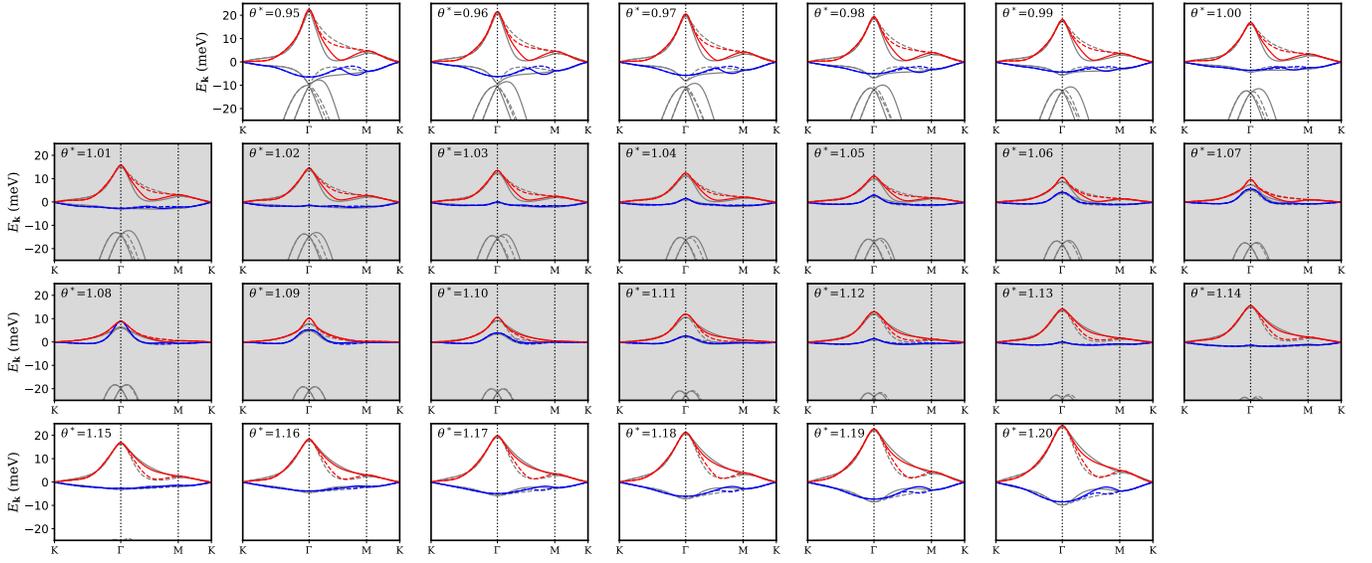

FIG. 2. Low-energy bands of the effective hamiltonian for angles $\theta^*$ between 0.95° and 1.20° calculated using the fit parameters described in the main text (red and blue), compared to the bands from the $\mathbf{k}\cdot\mathbf{p}$ model (gray). A light gray background indicates angles that lie within the magic range.

Even for energies $E \approx 0$, $\mathcal{H}_\mathbf{k}^{\text{eff}}$ has non-trivial momentum dependence due to the functions $f_i(\mathbf{k})$ that enter non-linearly in the second term of Eq. 1. However, near the K/K' valleys the expressions become simpler because $f_1$ and $f_3$ vanish,

$$f_1(\mathbf{K}) = 0, \quad f_2(\mathbf{K}) = -\frac{3}{2}, \quad f_3(\mathbf{K}) = 0, \tag{2}$$

so the hamiltonian takes the form

$$\mathcal{H}(\mathbf{K}) = \begin{pmatrix} t_0^* - \frac{3}{2}t_2^* & 0 & t_0^{\text{int}} - \frac{3}{2}t_2^{\text{int}} & 0 \\ 0 & t_0^* - \frac{3}{2}t_2^* & 0 & t_0^{\text{int}} - \frac{3}{2}t_2^{\text{int}} \\ t_0^{\text{int}} - \frac{3}{2}t_2^{\text{int}} & 0 & t_0^\triangle - \frac{3}{2}t_2^\triangle & 0 \\ 0 & t_0^{\text{int}} - \frac{3}{2}t_2^{\text{int}} & 0 & t_0^\triangle - \frac{3}{2}t_2^\triangle \end{pmatrix} \equiv \begin{pmatrix} A & 0 & B & 0 \\ 0 & A & 0 & B \\ B & 0 & C & 0 \\ 0 & B & 0 & C \end{pmatrix}, \tag{3}$$

and the effective hamiltonian is thus

$$\mathcal{H}^{\text{eff}}(\mathbf{K}) = \begin{pmatrix} A & 0 \\ 0 & A \end{pmatrix} - \begin{pmatrix} B & 0 \\ 0 & B \end{pmatrix}\begin{pmatrix} 1/C & 0 \\ 0 & 1/C \end{pmatrix}\begin{pmatrix} B & 0 \\ 0 & B \end{pmatrix} = \left( A - \frac{B^2}{C} \right)\mathbb{I}. \tag{4}$$

Thus, the low energy eigenvalues are doubly degenerate at K with a value

$$E(\mathbf{K}) = A - \left(\frac{B}{C}\right)B = \left(t_0^* - \frac{3}{2}t_2^*\right) - \frac{\left(t_0^{\text{int}} - \frac{3}{2}t_2^{\text{int}}\right)^2}{t_0^\triangle - \frac{3}{2}t_2^\triangle}, \tag{5}$$

which can be set to zero at charge neutrality by an overall energy shift in $t_0^*$ and $t_0^\triangle$. For small momenta near the high-symmetry K point we can expand the hamiltonian in powers of $k$:

$$\mathcal{H}(\mathbf{K}+k) = \begin{pmatrix} A & 0 & B & 0 \\ 0 & A & 0 & B \\ B & 0 & C & 0 \\ 0 & B & 0 & C \end{pmatrix} + \begin{pmatrix} 0 & a & 0 & b \\ a^* & 0 & b^* & 0 \\ 0 & b & 0 & c \\ b^* & 0 & c^* & 0 \end{pmatrix} k + \mathcal{O}(k^2) \tag{6}$$

where $k$ is the projection of $\mathbf{k}$ onto the K–Γ path, $a = \frac{\sqrt{3}}{2}a_0(-t_1^* + 2t_3^*)$, $b = \frac{\sqrt{3}}{2}a_0(-t_1^{\text{int}} + 2t_3^{\text{int}})$, $c = \frac{\sqrt{3}}{2}a_0(-t_1^\triangle + 2t_3^\triangle)$, and $a_0$ is the lattice constant of graphene. We assume all the hopping parameters are real except for $t_3^{\text{int}}$, which makes $b$ complex. The effective hamiltonian is then

$$\mathcal{H}^{\text{eff}}(0, \mathbf{K}+k) = \begin{bmatrix} A & ak \\ a^*k & A \end{bmatrix} - \frac{1}{C^2}\begin{bmatrix} B^2C & (2BCb - B^2c)k \\ (2BCb^* - B^2c^*)k & B^2C \end{bmatrix} + \mathcal{O}(k^2), \tag{7}$$

which has has eigenvalues $E = A - \left(\frac{B}{C}\right) B \pm \hbar v_F k$, and Fermi velocity given by

$$v_F = \frac{1}{\hbar} \left| a - 2\left(\frac{B}{C}\right) b + \left(\frac{B}{C}\right)^2 c \right|. \tag{8}$$

## FLAT BAND DENSITY OF STATES

Fig. 3 shows a single set of flat bands and the associated density of states (DOS) calculated using the $\mathbf{k} \cdot \mathbf{p}$ model.

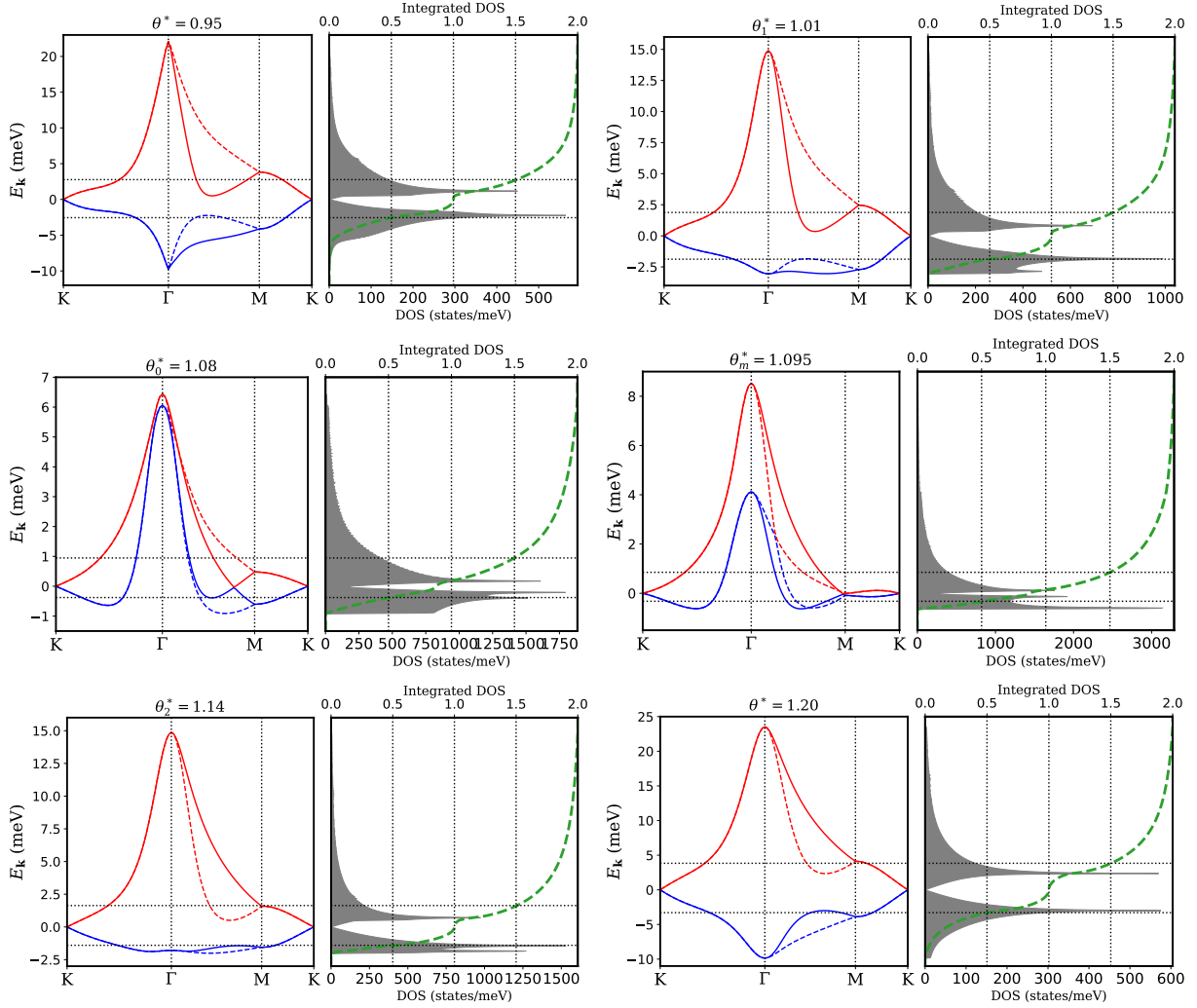

FIG. 3. Band structures and density of states (DOS) calculated using the $\mathbf{k} \cdot \mathbf{p}$ model for five selected twist angles $\theta^*$ between $0.95°$ and $1.20°$.



\end{document}